\documentclass[12pt]{article}
\usepackage{graphicx}

\def \bea{\begin{eqnarray}}
\def \beq{\begin{equation}}

\def \eea{\end{eqnarray}}
\def \eeq{\end{equation}}

\def \ket#1{| #1 \rangle}

\def \s{\sqrt{2}}
\def \st{\sqrt{3}}
\def \sx{\sqrt{6}}
\def \tth{\theta_{\ell \pi}}

\def \ve{\vec{\epsilon}}
\def \vp{\vec{p}}
\begin{document}
\rightline{EFI 04-30}
\rightline{hep-ph/0408334}
\bigskip

\centerline{\bf ANGULAR DISTRIBUTIONS IN $J/\psi (\rho^0,\omega)$ STATES
NEAR THRESHOLD
\footnote{To be submitted to Phys.\ Rev.\ D.}}
\bigskip
 
\centerline{Jonathan L. Rosner~\footnote{rosner@hep.uchicago.edu.}}
\centerline{\it Enrico Fermi Institute and Department of Physics}
\centerline{\it University of Chicago, 5640 S. Ellis Avenue, Chicago, IL 60637}
 
\begin{quote}
A resonance $X(3872)$, first observed in the decays $B \to K X$, has been
seen to decay to $J/\psi \pi^+ \pi^-$.  The $\pi^+ \pi^-$ mass spectrum peaks
near its kinematic upper limit, prompting speculation that the dipion system
may be in a $\rho^0$.  The decay $X(3872) \to J/\psi \omega$ also has been
observed.  The reaction $\gamma \gamma \to J/\psi \pi^+ \pi^-$ has been
studied.  Consequently, angular distributions in decays of $J/\psi
(\rho^0,\omega)$ states near threshold are of interest, and results are
presented.
\end{quote}

\noindent
PACS Categories: 12.39.Mk, 13.60.Le, 13.66.Bc, 14.40.Gx

\bigskip

\centerline{\bf I.  INTRODUCTION}
\bigskip

The spectrum of charmonium (bound states of a charmed quark $c$ and charmed
antiquark $\bar c$) follows some simple regularities, but the discovery of a
new resonance at 3872 MeV/$c^2$ \cite{Choi:2003ue,Acosta:2003zx,Abazov:2004kp,%
Aubert:2004ns} has posed some interesting puzzles \cite{Quigg:2004vf,%
Eichten:2004uh,Barnes:2003vb,Olsen:2004fp}.  This particle, initially observed
in $B \to K X(3872)$, was seen to decay to $J/\psi \pi^+ \pi^-$, prompting
speculation that it might be a charmonium level.  It is very narrow, and is
not seen decaying to $D \bar D$ although well above threshold
\cite{Chistov:2004}.  To account for this, it would have to have either a very
high spin $J \ge 3$ (thereby suppressing the $D \bar D$ by a sufficiently
strong centrifugal barrier \cite{Eichten:2004uh}), or to belong to the
unnatural spin-parity series $J^P = 0^-,~1^+,~2^-,\ldots$.

None of the $c \bar c$ assignments for $X(3872)$ is entirely satisfactory
\cite{Quigg:2004vf, Eichten:2004uh,Barnes:2003vb,Olsen:2004fp}.  An interesting
alternative \cite{Tornqvist:2003na,Close:2003sg} is that $X(3872)$ could be a
``molecular charmonium'' \cite{DeRujula:1976qd} state, since it lies very close
to the sum of the $D^0$ and $D^{*0}$ masses.  The peaking of the $\pi^+ \pi^-$
mass spectrum near the upper kinematic limit prompted the suggestion (see,
e.g., \cite{Choi:2003ue}) that the $\pi^+ \pi^-$ is in a $\rho^0$.  A bound
state of $D^0 \overline{D}^{*0} \pm {\rm c.c.}$ could be an admixture of $I=0$
and $I=1$ final states and thus capable of decaying to both $J/\psi \rho^0$ and
to $J/\psi \omega$ \cite{Swanson:2003tb}.  The calculation of Ref.\
\cite{Tornqvist:2003na} favors states with $J^P = 0^-$ and $1^+$, while that of
Ref.\ \cite{Swanson:2003tb} favors $1^+$.  If the state is decaying to $J/\psi
\rho^0$, its charge-conjugation eigenvalue must be $C = +$.  The Belle
Collaboration recently reported observation of the decay $X(3872) \to J/\psi
\omega$ at the expected rate for a ($D^0 \overline{D}^{*0} + {\rm c.c.}$)
S-wave bound state \cite{Abe:2004sd}.  The $J^P = 1^+$ assignment is highly
favored since the decays to $J/\psi \rho^0$ and are seen to occur near their
kinematic boundaries \cite{Voloshin:2004mh}.  Nonetheless it is of interest to
confirm this assignment by the study of angular distributions in the final
state.

A search for $X(3872)$ was undertaken in CLEO-III data \cite{Metreveli:2004}
for states produced in $\gamma \gamma$ fusion (with $C = +$ and $J \ne 1$) and
those produced in the radiative return reaction $e^+ e^- \to X(3872) + \gamma$
(with $J^{PC} = 1^{--}$ coupling to a virtual photon).  While the results of
this search (and one by the BES Collaboration for states produced in
radiative return \cite{Yuan:2003yz}) were negative, their interpretation
depends on the angular distribution in the final $J/\psi \pi^+ \pi^-$
products.  Under some circumstances the final states produced in $\gamma
\gamma$ reactions could consist of $J/\psi \rho^0$.  Many reactions in
which pairs of photons produce pairs of vector mesons have a peak near
threshold, so whether or not $\gamma \gamma \to J/\psi \rho^0$ is associated
with $X(3872)$, the behavior of its partial waves near threshold is of
interest \cite{Rosner:2004nh}.

The present paper gives angular distributions for
states decaying to $J/\psi (\to \ell^+ \ell^-) \rho^0 (\to \pi^+ \pi^-)$ near
threshold, using the transversity analysis developed for decays of $B$ mesons
to pairs of vector mesons \cite{Dunietz:1990cj,Dighe:1995pd} and other
appropriate methods.  These results can help to distinguish among some
hypotheses for the nature of the $X(3872)$ and to interpret $\gamma \gamma \to
J/\psi \pi^+ \pi^-$ reactions near threshold.  Results may be applied to
decays $J/\psi (\to \ell^+ \ell^-) \omega (\to \pi^+ \pi^- \pi^0)$ by
replacing a unit vector in the $\pi^+$ direction in the $\rho$ center-of-mass
system (c.m.s.) by the normal to the plane defined by $\pi^+ \pi^- \pi^0$ in
the $\omega$ rest frame.

The quantum number assignments to be considered are enumerated in Sec.\ II,
while the transversity and other formalisms are described in Sec.\ III.
Spinless cases are treated in Sec.\ IV, while decays of $J=1$ and $J=2$
positive-parity states are discussed in Secs.\ V and VI.  Section VII
concludes.  Appendices are devoted to angular distributions in vector meson
decays to $\pi^+ \pi^-$, $\pi^+ \pi^- \pi^0$, and $\ell^+ \ell^-$ and to an
elementary derivation of some angular correlations using Clebsch-Gordan
coefficients.
\bigskip

\centerline{\bf II. ENUMERATION OF $J^{PC}$ ASSIGNMENTS}
\bigskip

The states of $J/\psi$ ($J^{PC} = 1^{--}$) and $\rho^0$ ($J^{PC} = 1^{--}$)
near threshold may be divided into those of positive and negative
parity.  If the lowest possible relative orbital angular
momentum $L_{J/\psi,\rho}$ dominates, one must consider the states
in Table \ref{tab:states}.  Here $S_{J/\psi,\rho}$ denotes the spin of the
coupled $J/\psi$ and $\rho$ before consideration of their relative orbital
angular momentum.

\begin{table}
\caption{$J^{PC}$ values for states of $J/\psi \rho^0$ with definite parity
near threshold.
\label{tab:states}}
\begin{center}
\begin{tabular}{c c c} \hline \hline
$S_{J/\psi,\rho}$ & $L_{J/\psi,\rho}=0$ & $L_{J/\psi,\rho}=1$ \\
\hline
0          & $0^{++}$ & $1^{-+}$ \\
1          & $1^{++}$ & $0^{-+},~1^{-+},~2^{-+}$ \\
2          & $2^{++}$ & $1^{-+},~2^{-+},~3^{-+}$ \\ \hline \hline
\end{tabular}
\end{center}
\end{table}

For present purposes it is enough to consider a subset of the states
in Table \ref{tab:states}.  An S-wave bound state of ($D^0 \bar D^{*0} +
{\rm c.c.}$) must have $J^{PC} = 1^{++}$, while P-wave bound states can have
$J^{PC} = 0^{-+},~1^{-+},~2^{-+}$.  Of these only the first was proposed in
Ref.\ \cite{Tornqvist:2003na} as possibly bound.  The states accessible in
$\gamma \gamma$ collisions include those with $J^{PC} = 0^{\pm +}$,
$2^{\pm +}$, and $3^{-+}$; Yang's Theorem \cite{Yang:1950rg} forbids formation
of $J=1$ states.  In view of the complexity of the $2^{-+}$ and $3^{-+}$ 
states, they will be ignored.  Henceforth the discussion will concentrate on
the possibilities $0^{\pm +}$, $1^{++}$, and $2^{++}$.

\bigskip
\centerline{\bf III.  GENERAL FORMALISM}
\bigskip

A transversity analysis is helpful in analyzing decays to $J/\psi$ and another
vector meson $V$ when the $J/\psi$ decays to a lepton pair $\ell^+ \ell^-$ and
the vector meson decays to a pair of pseudoscalars $P^+ P^-$
\cite{Dunietz:1990cj,Dighe:1995pd}.  In the $P^+ P^-$ rest frame,
the $x$ axis is defined as the negative of the unit vector pointing in the
direction of the $J/\psi$.  The $P^+ P^-$ system is assumed to lie in the
$x$-$y$ plane, with $P^+$ making an angle $\psi$ with the $x$ axis ($0 \le \psi
\le \pi$).

The $z$ axis is taken in the $J/\psi$ rest frame perpendicular to the plane
containing the $P^+ P^-$ pair, using a right-handed coordinate system.  In
this frame the unit vector $\hat n(\ell^+)$ along the direction of the
positive lepton has coordinates $(n_x,n_y,n_z) = (\sin \theta \cos \varphi,
\sin \theta \sin \varphi, \cos \theta)$, thus defining $\theta$ and $\varphi$.
Results in the transversity basis will be presented for states with $J \le 1$.
For $J=2$ positive-parity states some simpler variables will be used.

In analyzing decays involving positive-parity $J/\psi \rho^0$ states near
threshold, for which $L_{J/\psi,\rho^0} = 0$ may be assumed, the rest frames of
the $J/\psi$ and $\rho^0$ may be assumed to coincide. One may then consider the
angle $\tth$ between the $\ell^+$ in $J/\psi \to \ell^+ \ell^-$ and the $\pi^+$
in $\rho^0 \to \pi^+ \pi^-$.  For such states produced in $\gamma \gamma$
collisions, angles $\theta_\ell$ and $\theta_{\pi^+}$ of $\ell^+$ and $\pi^+$
with respect to the photon
axis (in the $\gamma \gamma$ center of mass) also are useful.

\bigskip
\centerline{\bf IV.  STATES WITH $J^{PC} = 0^{\pm +}$}
\bigskip

The decays of spinless states to two vector mesons were discussed in Ref.\
\cite{Dighe:1995pd}.  In terms of amplitudes $A_0$, $A_\parallel$, and
$A_\perp$ describing longitudinal, parallel transverse, and perpendicular
transverse polarizations of the vector mesons, the differential distribution
may be written
$$
\frac{d^3 \Gamma [X \to (\ell^+\ell^-)_{J/\psi} (\pi^+ \pi^-)_{\rho}]}
{d \cos \theta~d \varphi~d \cos \psi}
= \frac{9}{32 \pi} [2 |A_0|^2 \cos^2 \psi (1 - \sin^2 \theta \cos^2 \varphi)
$$
$$
+ \sin^2 \psi \{ |A_\parallel|^2  (1 - \sin^2 \theta \sin^2 \varphi)
+ |A_\perp|^2 \sin^2 \theta - {\rm Im}(A_\parallel^* A_\perp)
\sin 2 \theta \sin \varphi \}~~~
$$
\beq \label{eqn:threeangle}
+\frac{1}{\sqrt{2}}\sin 2 \psi \{{\rm Re}(A_0^* A_\parallel) \sin^2 \theta
\sin 2 \varphi + {\rm Im}(A_0^* A_\perp) \sin 2 \theta \cos \varphi \} ]~~~.
\eeq
In terms of partial-wave amplitudes $S,~P,~D$ corresponding to $L=0,~1,~2$
between the vector mesons, one has
\beq
A_0 = - \sqrt{\frac{1}{3}} S + \sqrt{\frac{2}{3}} D~~,~~~
A_\parallel = \sqrt{\frac{2}{3}} S + \sqrt{\frac{1}{3}} D~~~,~~~
A_\perp = P~~~.
\eeq
The normalization has been chosen in such a way that the total width is
given by
\beq
\Gamma = |A_0|^2 + |A_\parallel|^2 + |A_\perp|^2 =
 |S|^2 + |P|^2 + |D|^2~~~.
\eeq
Specializing to the case of $0^{++}$ decaying via a pure S-wave, one finds
\begin{equation} \label{eqn:trans0}
\frac{1}{\Gamma} \frac{d^3 \Gamma}{d \cos \theta~d \varphi~d \cos \psi} =
\frac{3}{16 \pi} [1 - \sin^2 \theta \cos^2 (\psi - \varphi)]~~~.
\end{equation}

The dependence on $\psi - \varphi$ looks puzzling at first sight, since $\psi$
is a polar angle while $\varphi$ is an azimuthal angle.  However, $\psi$ is
defined with respect to the $x$ axis, while $\varphi$ is an azimuthal angle in
the $x$-$y$ plane.

The angular distribution is independent of $\psi$ or $\varphi$ when $\theta
=0$ or $\pi$.  In that case the $\ell^+ \ell^-$ and $\pi^+ \pi^-$ axes are
perpendicular to one another, and rotational invariance for the spinless
initial state guarantees that the angular distribution should not depend on
overall orientation.  The distribution vanishes when $\theta = \pi/2$ and
$\psi = \varphi$; in that case the $\ell^+ \ell^-$ and $\pi^+ \pi^-$ axes are
parallel to one another.  The $\ell^+ \ell^-$ state must always have
helicity $\pm 1$ along its axis, and hence cannot couple to a collinear
$\pi^+ \pi^-$ state if the total angular momentum is to vanish.  This last
property also allows one to derive a simple expression for the correlation
between the $\ell^+ \ell^-$ and $\pi^+ \pi^-$ axes in $0^{++}$ decay:
\beq \label{eqn:corr0}
\left. \frac{1}{\Gamma}\frac{d \Gamma}{d \cos \tth} \right|_{J^{PC} = 0^{++}}
 = \frac{3}{4} \sin^2 \tth~~~.
\eeq
This result also follows from Eq.\ (\ref{eqn:trans0}) by a simple
transformation of coordinate axes and integration over two of the three
variables.  Let the $\pi^+$ lie along the $+x$ axis (setting $\psi = 0$)
and note that $\sin \theta \cos \varphi$ is the cosine of the angle between
the $\ell^+$ and $\pi^+$.  An alternative derivation via Clebsch-Gordan
coefficients is contained in Appendix C.

The result for $0^{-+}$ decaying by a pure P-wave is quite simple:
\begin{equation}
\frac{1}{\Gamma} \frac{d^3 \Gamma}{d \cos \theta~d \varphi~d \cos \psi} =
\frac{9}{32 \pi} \sin^2 \theta \sin^2 \psi
\end{equation}
The proportionality of this distribution to $\sin^2 \theta$ is a general
feature of transversity amplitudes for decay of a CP-odd spinless state
(see Ref.\ \cite{Dighe:1995pd}).  The vanishing of the distribution when
$\psi = 0$ or $\pi$ is a feature of the coupling for a pseudoscalar decaying
to two vectors; the dipion axis (in the $\pi \pi$ rest frame) cannot be
parallel to the $J/\psi$ recoil momentum in that frame.  One cannot write a
simple relation corresponding to Eq.\ (\ref{eqn:corr0}) since the direction of
the recoil momentum is important.  The matrix element actually vanishes for
zero recoil momentum, as it does for all negative-parity states decaying to
$J/\psi \rho^0$.  This, in fact, is an argument against negative parity for
the $X(3872)$, since a matrix element proportional to recoil momentum would
inevitably suppress both the highest $\pi^+ \pi^-$ masses in the decay
$X(3872) \to J/\psi \rho^0$ and the observed process $X(3872) \to J/\psi
\omega$ \cite{Voloshin:2004mh}.

For both $0^{++}$ and $0^{-+}$ states produced in $\gamma \gamma$ collisions
and decaying to $J/\psi (\to \ell^+ \ell^-) \rho^0 (\to \pi^+ \pi^-)$,
individual distributions of lepton or pion pairs do not show any dependence
on the angles $\theta_\ell$ or $\theta_\pi$ with respect to the $\gamma \gamma$
axis.  This is a feature of the spinless nature of the initial state.

\bigskip
\centerline{\bf V.  STATES WITH $J^{PC} = 1^{++}$}
\bigskip

The angular distributions in a decay $X(1^{++}) \to J/\psi
\rho$ are best worked out using a Cartesian basis for the polarizations of
all three vector mesons.  The interaction Lagrangian is of the form
(in a nonrelativistic basis, which is satisfactory for an S-wave decay with
negligible recoil momentum)
\beq \label{eqn:triple}
{\cal L}_{\rm int} \propto \ve_X \cdot \ve_{J/\psi}^*
 \times \ve_\rho^* + {\rm c.c.}
\eeq
This simplifies calculations considerably.

It is convenient to neglect the relative momentum between the $J/\psi$
and the $\rho^0$ and to assume they (and the $X$) are in the same c.m.s.  Let
the direction in which the $X$ was boosted to reach its c.m.s. define the
$x$-axis, and let the $\pi^+$ lie in the $x$-$y$ plane, making an angle $\chi$
with the $x$-axis.  Thus a unit vector along the direction of the $\pi^+$ is
$\hat n(\pi^+) = \hat x \cos \chi + \hat y \sin \chi$, where $\hat x$ and $\hat
y$ are unit vectors along the $x$ and $y$ axes.  The polarization vector of
the $\rho$ must have this same form:  $\ve_\rho = \hat x \cos \chi +
\hat y \sin \chi$.  The $z$ axis is defined with respect to $\hat x$ and $\hat
y$ by a right-handed coordinate system.  Angles in polar coordinates are
defined so that a unit vector $\hat n$ with arbitrary direction has coordinates
\beq
\hat n = \hat x \sin \theta \cos \phi + \hat y \sin \theta \sin \phi
 + \hat z \cos \theta~~~.
\eeq

For $J(X)=1$, only two polarization states need be considered unless one is
interested in parity-violating processes.  These two polarizations in the frame
of interest are either along the $X$ boost direction, or
perpendicular to that direction, where one must sum over both such
perpendicular directions after squaring matrix elements.  The corresponding
amplitudes may be denoted $B_0$ and $B_T$, respectively.  The sum over two
transverse polarizations is equivalent to summing over two amplitudes
corresponding to helicity $\pm 1$, which must be equal if $X$ is produced in
parity-conserving processes.  Although $B$ decays do not necessarily conserve
parity, the effect of summing over $B$ and $\overline{B}$ decays is to
populate states of helicity $\pm 1$ equally.

For $X$ polarized along its boost direction, $\ve_X = \hat x$.  In
view of the coupling (\ref{eqn:triple}), only the $\hat y \sin \chi$ component
of $\ve_\rho$ can contribute, and the $J/\psi$ must be polarized along $\hat z$
(leading to a lepton distribution proportional to $\sin^2 \theta$).  Then the
full angular distribution of lepton pairs for this polarization state must be
proportional to $\sin^2 \theta \sin^2 \chi$.

For $X$ polarized transversely there are two contributions.  When $\ve_X = \hat
y$, only the $\hat x \cos \chi$ component of the $\rho$ polarization
contributes.  Again, the $J/\psi$ is polarized along $\hat z$ leading to a
lepton distribution $\sim \sin^2 \theta$.  Thus the angular distribution for
this polarization state is proportional to $\sin^2 \theta \cos^2 \chi$.

When $\ve_X = \hat z$,
the $J/\psi$ polarization vector is perpendicular to it and to $\ve_\rho$, so
$\ve_{J/\psi} = - \hat x \sin \chi + \hat y \cos \chi$.  The lepton
distribution for such a state (see Appendix B) is proportional to
$1 - \sin^2 \theta \sin^2(\chi - \phi)$.

The overall differential distribution, with longitudinal (0) and transverse (T)
amplitudes normalized so that $\Gamma = |B_0|^2 + 2 |B_T|^2$, is
$$
\frac{d^3 \Gamma}{d \cos \theta~d \phi~d \cos \chi} = \frac{9}
{32 \pi} \{ |B_0|^2 \sin^2 \theta \sin^2 \chi
$$
\beq
+ |B_T|^2 [ \sin^2 \theta \cos^2 \chi + 1 - \sin^2 \theta \sin^2 (\chi - \phi)
] \}~~~.
\eeq
If $X$ is unpolarized, corresponding to $|B_0| = |B_T|$, the result becomes
\beq
\frac{d^3 \Gamma}{d \cos \theta~d \phi~d \cos \chi} = \frac{9}
{32 \pi} |B_0|^2 [ 1 + \sin^2 \theta \cos^2(\chi - \phi) ]~~~
\eeq
This can be expressed as the simpler form
\beq
\left. \frac{1}{\Gamma}\frac{d \Gamma}{d \cos \tth} \right|_{J^{PC} = 1^{++}}
 = \frac{3}{8}(1 + \cos^2 \tth)~~~,
\eeq 
for which an alternative derivation is given in Appendix C.
% \newpage

\bigskip
\centerline{\bf VI.  STATES WITH $J^{PC} = 2^{++}$}
\bigskip

This assignment is probably not so relevant for the $X(3872)$, since there
are several arguments against it, including the absence of the decay
$X(3872) \to D \bar D$ which would be possible for a $2^{++}$ particle.
However, it is relevant for states of $J/\psi \rho^0$ such as can be produced
near threshold in $\gamma \gamma$ collisions, as studied in Ref.\
\cite{Metreveli:2004} and discussed in Ref.\ \cite{Rosner:2004nh}.

Two independent helicity states of a $2^{++}$ particle produced in $\gamma
\gamma$ collisions are possible:  $J_z = \pm 2$ and $J_z = 0$.  Assuming
the decay $2^{++} \to J/\psi \rho^0$ takes place with zero orbital angular
momentum and that the rest frames of $J/\psi$ and $\rho^0$ coincide, it is
a simple matter of Clebsch-Gordan coefficients to calculate the individual
distributions of leptons and pions with respect to the $\gamma \gamma$ axis.
For instance, when $J_z = \pm 2$, the individual vector mesons must have
$J_z = \pm 1$, and one should see the characteristic angular distributions
\beq
\frac{1}{\Gamma}\frac{d \Gamma}{d \cos \theta_\ell} = \frac{3}{8} (1 +
\cos^2 \theta_\ell)~~,~~~
\frac{1}{\Gamma}\frac{d \Gamma}{d \cos \theta_\pi} = \frac{3}{4} \sin^2
\theta_\pi~~~.
\eeq
The state with $J_z = 0$ is composed 1/3 of the time of vector mesons
with $J_z = \pm 1$ and 2/3 of the time of vector mesons with $J_z = 0$
(see Appendix C).
One then finds for the lepton angular distribution with respect to the beam
$$
\frac{1}{\Gamma}\frac{d \Gamma}{d \cos \theta_\ell} = \frac{1}{3} \left[
\frac{3}{8} (1 + \cos^2 \theta_\ell) \right ] + \frac{2}{3} \left[
\frac{3}{4} \sin^2 \theta_\ell \right ]
$$
\beq
= \frac{1}{8} \left( 5 - 3 \cos^2 \theta_\ell \right)
\eeq
and for the pion angular distribution
$$
\frac{1}{\Gamma}\frac{d \Gamma}{d \cos \theta_\ell} = \frac{1}{3} \left[
\frac{3}{4} \sin^2 \theta_\pi \right ] + \frac{2}{3} \left[
\frac{3}{2} \cos^2 \theta_\pi \right ]
$$
\beq
= \frac{1}{4} \left( 1 + 3 \cos^2 \theta_\pi \right)~~~.
\eeq
Putting these results together and defining normalized helicity amplitudes
$A_{\pm 2}$ and $A_0$ in such a way that $|A_0|^2 + 2|A_{\pm 2}|^2 = 1$, one
has
\beq
\frac{d \Gamma}{d \cos \theta_\ell} = \frac{1}{8}
\left[ 6|A_{\pm 2}|^2 \left( 1 + \cos^2 \theta_\ell \right)
 + |A_0|^2 \left( 5 - 3 \cos^2 \theta_\ell \right) \right]~~~,
\eeq
\beq
\frac{d \Gamma}{d \cos \theta_\pi} = \frac{1}{4}
\left[ 6|A_{\pm 2}|^2 \left( 1 - \cos^2 \theta_\pi \right)
 + |A_0|^2 \left( 1 + 3 \cos^2 \theta_\pi \right) \right]~~~.
\eeq

The correlation between $\ell$ and $\pi$ directions is independent of
the polarization of the $2^{++}$ state.  It is worked out in Appendix C
and is given by
\beq
\left. \frac{1}{\Gamma} \frac{d \Gamma}{d \cos \tth} \right|_{J^{PC}=2^{++}} =
\frac{3}{40} \left( 7 - \cos^2 \tth \right)
\eeq
% \newpage
\bigskip

\centerline{\bf VII.  CONCLUSIONS}
\bigskip

Several angular distributions for decays of $J/\psi \rho^0$ and $J/\psi \omega$
states near threshold have been given.  These apply both to the $X(3872)$,
whose favored quantum numbers of $J^{PC}$ are $1^{++}$ if it is indeed a
molecule of $D^0$ and $\overline{D}^{*0}$ \cite{Tornqvist:2003na,Close:2003sg,%
Voloshin:2004mh}, and to states formed near threshold by photon-photon
collisions.  A number of such states for lighter-quark systems have turned
out to have $J^{PC} = 2^{++}$ \cite{Rosner:2004nh}.  It will be interesting to
apply the present methods to the few events reported in Ref.\
\cite{Metreveli:2004} and to potentially larger samples available from
asymmetric $e^+ e^-$ colliders.
\bigskip

\centerline{\bf ACKNOWLEDGMENTS}
\bigskip

I thank Steve Olsen and Pete Zweber for discussions, and Maury Tigner for
extending the hospitality of the Laboratory for Elementary-Particle Physics at
Cornell during part of this research.  This work
was supported in part by the United States Department of Energy through Grant
No.\ DE FG02 90ER40560 and in part by the John Simon Guggenheim Memorial
Foundation.
\bigskip

\centerline{\bf APPENDIX A:  DECAYS OF POLARIZED $\rho$, $\omega$}
\bigskip

In the c.m.s. of the $\rho$, its coupling to two pions is of the form
\beq
{\cal L}_{\rm int} \propto \ve_\rho \cdot [\vp(\pi^+) - \vp(\pi^-)]~~~.
\eeq
Thus, a $\rho$ linearly polarized along the $\hat z$ direction ($\ve_\rho =
\hat z$ or $J_z^{\rho} = 0$) will give rise to a pion angular distribution
$\sim \cos^2 \theta$, where $\theta$ is the polar angle of the direction of
either pion, while a $\rho$ with $\ve_\rho = \mp (\hat x \pm i \hat y)/\s$
($J_z^{\rho} = \pm 1$) will have an angular distribution $\sim \sin^2 \theta/2$
(in the same normalization).  Adopting a normalization for which a distribution
$W(\cos \theta_\pi)$ obeys
\beq
\int_{-1}^{1} d \cos \theta_\pi W(\cos \theta_\pi) = 1~~~,
\eeq
one has
\beq
W_{J_z = 0}^\rho(\cos \theta_\pi) = \frac{3}{2} \cos^2 \theta_\pi~~,~~~
W_{J_z = \pm 1}^\rho(\cos \theta_\pi) = \frac{3}{4} \sin^2 \theta_\pi~~,~~~
\eeq
The coupling of an $\omega$ in its rest frame to $\pi^+ \pi^- \pi^0$ is of the
form ${\cal L}_{\rm int} \propto \ve_\omega \cdot\hat n$, where $\hat n$ is the
normal to the plane of decay defined by the three pions in the $\omega$ rest
frame.  Thus all results for $\rho^0 \to \pi^+ \pi^-$ may be transcribed
for $\omega$ decays by replacing a unit vector in the direction of the
$\pi^+$ in $\rho$ decay by the normal to the $\omega$ decay plane.
\bigskip

\centerline{\bf APPENDIX B:  DECAYS OF POLARIZED $J/\psi$}
\bigskip

The decay of a vector meson such as $J/\psi$ to $\ell^+(p_+) \ell^-(p_-)$
involves an interaction proportional to $\epsilon_\mu \bar u(p_-) \gamma^\mu
v(p_+)$, where $u$ and $v$ are Dirac spinors.  Performing the sum over
lepton spins, and denoting the initial $J/\psi$ spin configuration by a
density matrix $\rho_{J/\psi}$, one finds that a general angular distribution
in the $J/\psi$ c.m.s.  for any $\rho_{J/\psi}$ is proportional to
Tr($\rho_{J/\psi} L$), where $L$ is a $3 \times 3$ matrix: $L_{ij} =
\delta_{ij} - n^i n^j$, with $n^i$ denoting the Cartesian coordinates of the
$\ell^+$ momentum.  In spherical polar coordinates, one has
\beq
L = \left[ \begin{array}{c c c}
 1 - \sin^2 \theta \cos^2 \phi & - \sin^2 \theta \sin \phi \cos \phi &
   - \sin \theta \cos \theta \cos \phi \\
 - \sin^2 \theta \sin \phi \cos \phi & 1 - \sin^2 \theta \sin^2 \phi &
   - \sin \theta \cos \theta \sin \phi \\
 - \sin \theta \cos \theta \cos \phi & - \sin \theta \cos \theta \sin \phi &
 1 - \cos^2 \theta \end{array} \right]~~~.
\eeq

The spin density matrix for a pure $J/\psi$ polarization state $\epsilon_i$
is $\rho_{ij} = \epsilon_i \epsilon^*_j$.  Thus, for example, $\rho(J_z = 0)
= {\rm diag}(0,0,1)$ and the corresponding angular distribution for lepton
pairs is proportional to $1 - \cos^2 \theta = \sin^2 \theta$.  A $J/\psi$ is
produced in $e^+ e^-$ collisions in a mixed state, half $J_z = 1$ and half
$J_z = -1$, corresponding to the density matrix $\rho = (1/2){\rm diag}
(1,1,0)$, so the corresponding angular distribution for final leptons is
proportional to $(1/2)(2 - \sin^2 \theta) = (1/2)(1 + \cos^2 \theta)$.  The
suitably normalized forms are
\beq
W_{J_z^{J/\psi} = 0} = (3/4) \sin^2 \theta~~,~~~
W_{J_z^{J/\psi} = \pm 1~({\rm mixed})} = (3/8)(1 + \cos^2 \theta)~~~.
\eeq
\bigskip

\centerline{\bf APPENDIX C:  CORRELATIONS IN $\tth$}
\bigskip

The decays discussed here are of the form $X \to J/\psi (\to \ell^+ \ell^-)
\rho^0 (\to \pi^+ \pi^-)$ when the $J/\psi$ and $\rho$ are in a relative
S-wave and when their relative momentum may be neglected.  The angle $\tth$
is defined as that between the $\ell^+$ and $\pi^+$ in the common rest frame
of $J/\psi$ and $\rho^0$.

Let the $\pi^+$ direction define the $z$-axis.  States $|J,J_z)$ of definite
total $J = 0,~,1~,2$ and definite $J_z$ may now be decomposed into linear
combinations of substates $\ket{J_z^{J/\psi}, J_z^{\rho}}$.  Since the
$\pi^+$ has been taken along the $z$ axis, one is only concerned with
cases with $J_z^{\rho} = 0$.  As shown in Appendix B, states with definite
$J_z^{J/\psi}$ lead to distributions
\beq
W_{J_z^{J/\psi} = \pm 1}(\cos \tth) = \frac{3}{8} \left( 1 + \cos^2 \tth
\right)~~,~~~ W_{J_z^{J/\psi} = 0}(\cos \tth) = \frac{3}{4} \sin^2 \tth~~~,
\eeq
where these distributions are normalized according to
\beq
\int_{-1}^{1} d \cos \tth W(\cos \tth) = 1~~~.
\eeq

One must sum over all the $J_z$ values and divide by $2J + 1$ (the number of
$J_z$ values) to obtain the final result.  One must also multiply by 3 to
take account of the fact that a specific polarization state $J_z = 0$ of
the $\rho^0$ has been chosen.  As an example, a state with $J=J_z = 0$ has the
decomposition
\beq
|0,0) = \frac{1}{\st} \left[ \ket{1,-1} - \ket{0,0} + \ket{-1,1} \right]~~~.
\eeq
Only the substate $\ket{0,0}$ contributes, with weight (the square of the
Clebsch-Gordan coefficient) equal to 1/3.  One thus finds
\beq
\left. \frac{1}{\Gamma} \frac{d \Gamma}{d \cos \tth} \right|_{J=0} = 3 \cdot
\frac{1}{3} \cdot \frac{3}{4} \sin^2 \tth = \frac{3}{4} \sin^2 \tth~~~.
\eeq

For $J=1$ one uses the decomposition
\beq
|1 \pm 1 ) = \frac{1}{\s} \pm \left[ \ket{\pm 1, 0} - \ket{0, \pm 1} \right]~~,
\eeq
noting that $|1, 0)$ has no contribution from $\ket{0,0}$.  Then one finds
\beq
\left. \frac{1}{\Gamma} \frac{d \Gamma}{d \cos \tth} \right|_{J=1} = 3 \cdot
\frac{1}{3} \cdot \frac{1}{2} \cdot 2 \cdot \frac{3}{8} (1 + \cos^2 \tth) =
\frac{3}{8} (1 + \cos^2 \tth)~~~,
\eeq
where the factor of $1/3$ is $1/(2J+1)$, the factor of $1/2$ is the square of
the Clebsch-Gordan coefficient, and the factor of 2 corresponds to the two
substates $J_z = \pm 1$.  Finally, for $J=2$, one uses the decompositions
\beq
 |2,\pm 1) = \frac{1}{\s} \left[ \ket{\pm 1, 0} + \ket{0, \pm 1} \right]~~,
\eeq
\beq
 |2,0) = \frac{1}{\sx} \left[ \ket{1,-1} + 2\ket{0,0} + \ket{-1,1} \right]
\eeq
to obtain
\beq
\left. \frac{1}{\Gamma} \frac{d \Gamma}{d \cos \tth} \right|_{J=2} =
\frac{3}{40} \left( 7 - \cos^2 \tth \right)
\eeq
after a similar calculation.  (Note that the substates with $J_z = \pm 2$
do not involve $J_z^\rho = 0$ and hence do not contribute.)  The results
are summarized in Table \ref{tab:corr}.

\begin{table}
\caption{Normalized distributions $(1/\Gamma)d \Gamma / d \cos
\tth$ in cosine of the angle between the $\ell^+ \ell^-$ and
$\pi^+ \pi^-$ axes, for decays to S-wave states of $J/\psi \rho^0$ near
threshold.
\label{tab:corr}}
\begin{center}
\begin{tabular}{c c} \hline \hline
$J$ & Distribution \\ \hline
 0  & $(3/4)\sin^2 \tth$ \\
 1  & $(3/8)(1 + \cos^2 \tth)$ \\
 2  & $(3/40)(7 - \cos^2 \tth)$ \\ \hline \hline
\end{tabular}
\end{center}
\end{table}

% Journal and other miscellaneous abbreviations for references
% Phys. Rev. D format
\def \ajp#1#2#3{Am.\ J. Phys.\ {\bf#1}, #2 (#3)}
\def \apny#1#2#3{Ann.\ Phys.\ (N.Y.) {\bf#1}, #2 (#3)}
\def \app#1#2#3{Acta Phys.\ Polonica {\bf#1}, #2 (#3)}
\def \arnps#1#2#3{Ann.\ Rev.\ Nucl.\ Part.\ Sci.\ {\bf#1}, #2 (#3)}
\def \art{and references therein}
\def \cmts#1#2#3{Comments on Nucl.\ Part.\ Phys.\ {\bf#1}, #2 (#3)}
\def \cn{Collaboration}
\def \cp89{{\it CP Violation,} edited by C. Jarlskog (World Scientific,
Singapore, 1989)}
\def \efi{Enrico Fermi Institute Report No.\ }
\def \epjc#1#2#3{Eur.\ Phys.\ J. C {\bf#1}, #2 (#3)}
\def \f79{{\it Proceedings of the 1979 International Symposium on Lepton and
Photon Interactions at High Energies,} Fermilab, August 23-29, 1979, ed. by
T. B. W. Kirk and H. D. I. Abarbanel (Fermi National Accelerator Laboratory,
Batavia, IL, 1979}
\def \hb87{{\it Proceeding of the 1987 International Symposium on Lepton and
Photon Interactions at High Energies,} Hamburg, 1987, ed. by W. Bartel
and R. R\"uckl (Nucl.\ Phys.\ B, Proc.\ Suppl., vol.\ 3) (North-Holland,
Amsterdam, 1988)}
\def \ib{{\it ibid.}~}
\def \ibj#1#2#3{~{\bf#1}, #2 (#3)}
\def \ichep72{{\it Proceedings of the XVI International Conference on High
Energy Physics}, Chicago and Batavia, Illinois, Sept. 6 -- 13, 1972,
edited by J. D. Jackson, A. Roberts, and R. Donaldson (Fermilab, Batavia,
IL, 1972)}
\def \ijmpa#1#2#3{Int.\ J.\ Mod.\ Phys.\ A {\bf#1}, #2 (#3)}
\def \ite{{\it et al.}}
\def \jhep#1#2#3{JHEP {\bf#1}, #2 (#3)}
\def \jpb#1#2#3{J.\ Phys.\ B {\bf#1}, #2 (#3)}
\def \lg{{\it Proceedings of the XIXth International Symposium on
Lepton and Photon Interactions,} Stanford, California, August 9--14 1999,
edited by J. Jaros and M. Peskin (World Scientific, Singapore, 2000)}
\def \lkl87{{\it Selected Topics in Electroweak Interactions} (Proceedings of
the Second Lake Louise Institute on New Frontiers in Particle Physics, 15 --
21 February, 1987), edited by J. M. Cameron \ite~(World Scientific, Singapore,
1987)}
\def \kdvs#1#2#3{{Kong.\ Danske Vid.\ Selsk., Matt-fys.\ Medd.} {\bf #1},
No.\ #2 (#3)}
\def \ky85{{\it Proceedings of the International Symposium on Lepton and
Photon Interactions at High Energy,} Kyoto, Aug.~19-24, 1985, edited by M.
Konuma and K. Takahashi (Kyoto Univ., Kyoto, 1985)}
\def \mpla#1#2#3{Mod.\ Phys.\ Lett.\ A {\bf#1}, #2 (#3)}
\def \nat#1#2#3{Nature {\bf#1}, #2 (#3)}
\def \nc#1#2#3{Nuovo Cim.\ {\bf#1}, #2 (#3)}
\def \nima#1#2#3{Nucl.\ Instr.\ Meth. A {\bf#1}, #2 (#3)}
\def \np#1#2#3{Nucl.\ Phys.\ {\bf#1}, #2 (#3)}
\def \npbps#1#2#3{Nucl.\ Phys.\ B Proc.\ Suppl.\ {\bf#1}, #2 (#3)}
\def \os{XXX International Conference on High Energy Physics, Osaka, Japan,
July 27 -- August 2, 2000}
\def \PDG{Particle Data Group, K. Hagiwara \ite, \prd{66}{010001}{2002}}
\def \pisma#1#2#3#4{Pis'ma Zh.\ Eksp.\ Teor.\ Fiz.\ {\bf#1}, #2 (#3) [JETP
Lett.\ {\bf#1}, #4 (#3)]}
\def \pl#1#2#3{Phys.\ Lett.\ {\bf#1}, #2 (#3)}
\def \pla#1#2#3{Phys.\ Lett.\ A {\bf#1}, #2 (#3)}
\def \plb#1#2#3{Phys.\ Lett.\ B {\bf#1}, #2 (#3)}
\def \pr#1#2#3{Phys.\ Rev.\ {\bf#1}, #2 (#3)}
\def \prc#1#2#3{Phys.\ Rev.\ C {\bf#1}, #2 (#3)}
\def \prd#1#2#3{Phys.\ Rev.\ D {\bf#1}, #2 (#3)}
\def \prl#1#2#3{Phys.\ Rev.\ Lett.\ {\bf#1}, #2 (#3)}
\def \prp#1#2#3{Phys.\ Rep.\ {\bf#1}, #2 (#3)}
\def \ptp#1#2#3{Prog.\ Theor.\ Phys.\ {\bf#1}, #2 (#3)}
\def \rmp#1#2#3{Rev.\ Mod.\ Phys.\ {\bf#1}, #2 (#3)}
\def \rp#1{~~~~~\ldots\ldots{\rm rp~}{#1}~~~~~}
\def \rpp#1#2#3{Rep.\ Prog.\ Phys.\ {\bf#1}, #2 (#3)}
\def \sing{{\it Proceedings of the 25th International Conference on High Energy
Physics, Singapore, Aug. 2--8, 1990}, edited by. K. K. Phua and Y. Yamaguchi
(Southeast Asia Physics Association, 1991)}
\def \slc87{{\it Proceedings of the Salt Lake City Meeting} (Division of
Particles and Fields, American Physical Society, Salt Lake City, Utah, 1987),
ed. by C. DeTar and J. S. Ball (World Scientific, Singapore, 1987)}
\def \slac89{{\it Proceedings of the XIVth International Symposium on
Lepton and Photon Interactions,} Stanford, California, 1989, edited by M.
Riordan (World Scientific, Singapore, 1990)}
\def \smass82{{\it Proceedings of the 1982 DPF Summer Study on Elementary
Particle Physics and Future Facilities}, Snowmass, Colorado, edited by R.
Donaldson, R. Gustafson, and F. Paige (World Scientific, Singapore, 1982)}
\def \smass90{{\it Research Directions for the Decade} (Proceedings of the
1990 Summer Study on High Energy Physics, June 25--July 13, Snowmass, Colorado),
edited by E. L. Berger (World Scientific, Singapore, 1992)}
\def \tasi{{\it Testing the Standard Model} (Proceedings of the 1990
Theoretical Advanced Study Institute in Elementary Particle Physics, Boulder,
Colorado, 3--27 June, 1990), edited by M. Cveti\v{c} and P. Langacker
(World Scientific, Singapore, 1991)}
\def \yaf#1#2#3#4{Yad.\ Fiz.\ {\bf#1}, #2 (#3) [Sov.\ J.\ Nucl.\ Phys.\
{\bf #1}, #4 (#3)]}
\def \zhetf#1#2#3#4#5#6{Zh.\ Eksp.\ Teor.\ Fiz.\ {\bf #1}, #2 (#3) [Sov.\
Phys.\ - JETP {\bf #4}, #5 (#6)]}
\def \zpc#1#2#3{Zeit.\ Phys.\ C {\bf#1}, #2 (#3)}
\def \zpd#1#2#3{Zeit.\ Phys.\ D {\bf#1}, #2 (#3)}

\end{document}